\documentclass{ws-ijmpcs}
\usepackage[usenames]{color}
\usepackage{dcolumn}
\allowdisplaybreaks

\begin{document}

\markboth{Y. Ma \textit{et al.}}
{Skyrmions, half-skyrmions and nucleon mass in dense baryonic matter}

%
\catchline{}{}{}{}{}
%

\title{Skyrmions, half-skyrmions and nucleon mass in dense baryonic matter}


\author{Yong-Liang Ma}

\address{Department of Physics,  Nagoya University, Nagoya, 464-8602, Japan \\
ylma@hken.phys.nagoya-u.ac.jp}

\author{Masayasu Harada}

\address{Department of Physics,  Nagoya University, Nagoya, 464-8602, Japan \\
harada@hken.phys.nagoya-u.ac.jp}

\author{Hyun Kyu Lee}

\address{Department of Physics, Hanyang University, Seoul 133-791, Korea \\
hyunkyu@hanyang.ac.kr}

\author{Yongseok Oh}

\address{Department of Physics, Kyungpook National University, Daegu 702-701, Korea \\
Asia Pacific Center for Theoretical Physics, Pohang, Gyeongbuk 790-784, Korea \\
yohphy@knu.ac.kr}

\author{Mannque Rho}

\address{Department of Physics, Hanyang University, Seoul 133-791, Korea\\
Institut de Physique Th\'eorique, CEA Saclay, 91191 Gif-sur-Yvette c\'edex, France \\
mannque.rho@cea.fr}

\maketitle


\begin{abstract}

We explore the hadron properties in dense baryonic matter in a unified way by
using a Skyrme model constructed with an effective Lagrangian which includes
the $\rho$ and $\omega$ vector mesons as hidden gauge bosons and is valid up to
$O(p^4)$ in chiral expansion including the homogeneous Wess-Zumino terms.
With the two input values of pion decay constant and the lowest lying
vector meson mass which can be fixed in free space, all the other low energy constants in the effective Lagrangian are determined by their master formulas derived from holographic
QCD models, which allows us to study the baryonic matter properties with no additional
free parameters and thus without ambiguities.
We find that the $\omega$ field that figures in the homogeneous Wess-Zumino term
plays a crucial role in the skyrmion structure and its matter properties.
The most striking and intriguing observation is that the pion decay constant that
smoothly drops with increasing density in the Skyrmion phase stops decreasing at
$n_{1/2}^{}$ at which the skyrmions in medium fractionize into half-skyrmions and
remains nearly constant in the half-skyrmion phase.
In accordance with the large $N_c$ consideration, the baryon mass also stays
non-scaling in the half-skyrmion phase.
This feature is supported by the nuclear effective field theory with the parameters
of the Lagrangian scaling modified at the skyrmion--half-skyrmion phase transition.
Our exploration also uncovers the crucial role of the $\omega$ meson in
multi-baryon systems as well as in the structure of a single skyrmion.

\end{abstract}

\ccode{PACS numbers:
12.39.Dc ,  
12.39.Fe, 
21.65.-f,	
21.65.Jk  
}

\section{Introduction}

The physics of baryonic matter poses a challenge in nuclear and particle physics
and is crucial to understand the equation of state relevant for compact stars and the QCD phase structure at low temperature.
The difficulties in accessing compressed baryonic matter arise from the highly nonperturbative nature of strong interactions, lacking both experimental data and trustful theoretical tools. The situation is acerbated by that QCD lattice simulation suffers from the famous sign problem. As a possible way to access this difficult problem, we apply Skyrme's approach\cite{Skyrme62}, where the classical soliton solutions of the mesonic theory of QCD are interpreted as baryons in the sense of large $N_c$ limit.
This model was proposed in Ref.~\refcite{LPMRV03} as a natural framework to explore compressed baryonic matter by putting the skyrmions on the crystal lattice\cite{PV09}. This approach has the merit that allows one to study  in a unified way the properties of elementary baryon, many-baryon systems and medium-modified hadron properties.

In pursuance of Skyrme's seminal idea, we explored the nuclear
matter and in-medium hadron properties by the hidden local symmetry (HLS) Lagrangian expanded to the chiral order $O(p^4)$ including the lowest lying $\rho$ and $\omega$ vector mesons\cite{BKUYY85,BKY88,HY03a}.
Relegating details to Refs.~\refcite{LPMRV03,MOYHLPR12,MYOH12,MYOHLPR13,MHLOPR13}, we give a brief summary of the main ideas and the main results of our approach.
In comparison with what is available in the existing literature, our approach has the following distinctive features:

\begin{itemize}

\item
We use a chiral effective theory of mesons, which has a self-consistent chiral
counting mechanism, including pseudoscalar and lowest lying $\rho$ and $\omega$
vector mesons.
The full Lagrangian up to the next to leading order of the chiral counting including
the homogeneous Wess-Zumino (hWZ) terms  is applied.

\item
The low energy constants (LECs) of the effective Lagrangian are
self-consistently determined by a set of master formulas derived from a class of
holographic QCD (hQCD) models.
Therefore, we can fix their values without ambiguity with only two input values of the
pion decay constant $f_\pi$ and the vector meson mass $m_\rho$.

\item
Using this Lagrangian with the LECs determined as above, we explore the
single skyrmion, skyrmion matter, and medium-modified hadron properties.
The skyrmion matter is accessed by putting the skyrmions onto the face-centered
cubic (FCC) crystal.
By regarding mesons as fluctuations with respect to the skyrmion, we can study the
in-medium properties of hadrons.

\item
We include both the isovector $\rho$ meson and isoscalar $\omega$ meson
to explore the different roles of these mesons in nuclear matter.

\end{itemize}

Using the chiral order $O(p^4)$ HLS Lagrangian we found the followings.

\begin{itemize}

\item
Compared with the original Skyrme model that includes only pions, the existence
of the $\rho$ meson reduces the skyrmion mass and suppresses its radius.
However, the $\omega$ meson effect is exactly opposite to that of the $\rho$ meson,
i.e., with the existence of the $\omega$ the skyrmion mass is enhanced and
its radius increases.

\item The vector mesons affect the skyrmion matter properties notably.
The $\rho$ meson favors a high value for the half-skyrmion phase transition
density $n_{1/2}^{}$, while the $\omega$ prefers a smaller value of $n_{1/2}^{}$.
When the both effects are included $n_{1/2}^{}$ becomes similar to the normal
nuclear density $n_0^{}$.

\item In nuclear matter, $f_\pi^{\ast}$ decreases smoothly with increasing density
until the density reaches $n_{1/2}^{}$ and after that it becomes more or less a
nonzero constant.
Moreover, the in-medium nucleon mass $m_N^{\ast}$ also has a similar
density-dependence. These similarities are supported by the scaling of $m_N^{} \propto  f_\pi/e \sim (\sqrt{N_c})^2$ in the Skyrme model
where the Skyrme parameter $e$ is $O(1/\sqrt{N_c})$.

\end{itemize}

The features of our approach summarized above will be discussed in more detail
in the following manner.
We begin, in Sec.~\ref{sec:baryon}, with providing the HLS Lagrangian that is
the building block of our model for the structure and matter properties of skyrmions.
We also introduce the master formulas for the LECs based on hQCD.
Then the skyrmion properties are explored in this approach and the skyrmion
matter properties are studied by putting the skyrmions on the FCC crystal in
Sec.~\ref{sec:matter}.
Special emphasis is put on the role of vector mesons on the value of $n_{1/2}^{}$,
where the half-skyrmion phase appears.
In Sec.~\ref{sec:hadron} we discuss the hadron properties in nuclear medium, and
summarize the results and discussions in Sec.~\ref{sec:sum}.

\section{Baryon properties in hidden local symmetry}
\label{sec:baryon}

The HLS Lagrangian under consideration for investigating the structure and
matter properties of skyrmions has the symmetry group
$G_{\rm full} = [ \mbox{SU(2)}_L \times \mbox{SU(2)}_R ]_{\rm chiral} \times
[\mbox{U(2)}]_{\rm HLS}$.
Here, $[\mbox{U(2)}]_{\rm HLS}$ is adopted as the ``hidden local symmetry" to incorporate
the $\rho$ and $\omega$ vector mesons.
The basic building blocks of HLS Lagrangian are two 1-forms $\hat{\alpha}_{\parallel \mu}$
and $\hat{\alpha}_{\perp \mu}$ defined by
\begin{eqnarray}
\hat{\alpha}_{\parallel \mu}^{} &=&
\frac{1}{2i} (D_\mu^{} \xi_R^{} \xi_R^\dagger + D_\mu^{} \xi_L^{} \xi_L^\dagger), \qquad
\hat{\alpha}_{\perp \mu} =
\frac{1}{2i} (D_\mu^{} \xi_R^{} \xi_R^\dagger - D_\mu^{} \xi_L^{} \xi_L^\dagger),
\end{eqnarray}
with the chiral fields $\xi_L^{}$ and $\xi_R^{}$, which are expressed in the unitary gauge as
\begin{eqnarray}
\xi_L^\dagger &=& \xi_R^{} = e^{i \pi/2f_\pi} \equiv \xi
\qquad \mbox{with} \qquad
\pi = \bm{\pi} \cdot {\bm{\tau}}/{2},
\end{eqnarray}
where $\bm{\tau}$'s are the Pauli matrices. The covariant derivative is defined as
\begin{eqnarray}
D_\mu \xi_{R,L}^{} &=& (\partial_\mu - i V_\mu) \xi_{R,L}^{}
\end{eqnarray}
where $V_\mu = \frac{g}{2} \left( \omega_\mu + \rho_\mu \right)$ is the gauge boson
of the HLS with $ \rho_\mu = \bm{\rho}_\mu \cdot \bm{\tau}$.

Then the HLS Lagrangian adopted in this work can be written as
\begin{eqnarray}
\mathcal{L}_{\rm HLS} &=&
\mathcal{L}_{(2)} + \mathcal{L}_{(4)} + \mathcal{L}_{\rm anom} .
\label{eq:Lag_HLS}
\end{eqnarray}
The chiral order $O(p^2)$ term, $\mathcal{L}_{(2)}$, is constructed as\cite{BKUYY85,BKY88,HY03a}
\begin{eqnarray}
\mathcal{L}_{(2)} &=&
f_\pi^2 \,\mbox{Tr}\, \left( \hat{\alpha}_{\perp\mu}^{} \hat{\alpha}_{\perp}^{\mu} \right)
+ a f_\pi^2 \,\mbox{Tr}\, \left(\hat{\alpha}_{\parallel\mu}^{} \hat{\alpha}_{\parallel}^{\mu} \right)
- \frac{1}{2g^2} \mbox{Tr}\, \left( V_{\mu\nu} V^{\mu\nu} \right),
\end{eqnarray}
where $f_\pi$ is the pion decay constant, $a$ is the HLS parameter,
$g$ is the vector meson coupling constant, and the field-strength tensor of
the vector meson is $ V_{\mu\nu} = \partial_\mu V_\nu - \partial_\nu V_\mu - i [V_\mu,V_\nu]$.
The $O(p^4)$ Lagrangian at the leading order in $N_c$ is given by\cite{HY03a}
\begin{eqnarray}
\mathcal{L}_{(4)} & = & \mathcal{L}_{(4)y} + \mathcal{L}_{(4)z} , \nonumber\\
\mathcal{L}_{(4)y} &=&
y_1^{} \mbox{Tr} \Bigl[ \hat{\alpha}_{\perp\mu}^{} \hat{\alpha}_\perp^\mu
\hat{\alpha}_{\perp\nu}^{} \hat{\alpha}_\perp^\nu \Bigr]
+ y_2^{} \mbox{Tr} \Bigl[ \hat{\alpha}_{\perp\mu}^{} \hat{\alpha}_{\perp\nu}^{}
\hat{\alpha}^\mu_\perp \hat{\alpha}^\nu_\perp \Bigr] \nonumber\\
& & {} + y_3^{} \mbox{Tr}
\left[ \hat{\alpha}_{\parallel\mu}^{} \hat{\alpha}_\parallel^\mu
\hat{\alpha}_{\parallel\nu}^{} \hat{\alpha}_\parallel^\nu \right]
+ y_4^{} \mbox{Tr}
\left[ \hat{\alpha}_{\parallel\mu}^{} \hat{\alpha}_{\parallel\nu}^{}
\hat{\alpha}^\mu_\parallel \hat{\alpha}^\nu_\parallel \right]
\nonumber \\ && \mbox{}
+ y_5^{} \mbox{Tr}
\left[ \hat{\alpha}_{\perp\mu}^{} \hat{\alpha}_\perp^\mu
\hat{\alpha}_{\parallel\nu}^{} \hat{\alpha}_\parallel^\nu \right]
+ y_6^{} \mbox{Tr}
\left[ \hat{\alpha}_{\perp\mu}^{} \hat{\alpha}_{\perp\nu}^{}
\hat{\alpha}^\mu_\parallel \hat{\alpha}^\nu_\parallel \right]
+ y_7^{} \mbox{Tr}
\left[ \hat{\alpha}_{\perp\mu}^{} \hat{\alpha}_{\perp\nu}^{}
\hat{\alpha}^\nu_\parallel \hat{\alpha}^\mu_\parallel \right]
\nonumber \\ && \mbox{}
+ y_8^{} \left\{
\mbox{Tr} \left[ \hat{\alpha}_{\perp\mu}^{} \hat{\alpha}_\parallel^\mu
\hat{\alpha}_{\perp\nu}^{} \hat{\alpha}_\parallel^\nu \right]
+ \mbox{Tr} \left[ \hat{\alpha}_{\perp\mu}^{} \hat{\alpha}_{\parallel\nu}^{}
\hat{\alpha}_\perp^\nu \hat{\alpha}_\parallel^\mu \right] \right\}
+ y_9^{} \mbox{Tr}
\left[ \hat{\alpha}_{\perp\mu}^{} \hat{\alpha}_{\parallel\nu}^{}
\hat{\alpha}^\mu_\perp \hat{\alpha}^\nu_\parallel \right],\nonumber\\
\mathcal{L}_{(4)z} & = &
i z_4^{} \mbox{Tr}
\Bigl[ V_{\mu\nu} \hat{\alpha}_\perp^\mu \hat{\alpha}_\perp^\nu \Bigr]
+ i z_5^{} \mbox{Tr}
\left[ V_{\mu\nu} \hat{\alpha}_\parallel^\mu \hat{\alpha}_\parallel^\nu \right].
\end{eqnarray}
Finally, the anomalous-parity homogeneous WZ (hWZ) terms are written as
\begin{eqnarray}
\Gamma_{\rm hWZ} & = & \int d^4x \mathcal{L}_{\rm anom} = \frac{N_c}{16\pi^2}\int_{M^4}
\sum_{i=1}^3 c_i \mathcal{L}_i ,
\end{eqnarray}
where $M^4$ stands for the four-dimensional Minkowski space and
\begin{eqnarray}
&& \mathcal{L}_1 = i \, \mbox{Tr}\,
\bigl[ \hat{\alpha}_{\rm L}^3 \hat{\alpha}_{\rm R}^{}
 - \hat{\alpha}_{\rm R}^3 \hat{\alpha}_{\rm L}^{} \bigr], \quad
\mathcal{L}_2 = i \, \mbox{Tr}\,
\bigl[ \hat{\alpha}_{\rm L}^{} \hat{\alpha}_{\rm R}^{}
\hat{\alpha}_{\rm L}^{} \hat{\alpha}_{\rm R}^{} \bigr]  ,
\nonumber \\ &&
\mathcal{L}_3 = \mbox{Tr}\,
\bigl[ F_{\rm V} \left( \hat{\alpha}_{\rm L}^{} \hat{\alpha}_{\rm R}^{}
 - \hat{\alpha}_{\rm R}^{} \hat{\alpha}_{\rm L}^{} \right) \bigr] ,
\end{eqnarray}
in the 1-form and 2-form notations with
\begin{eqnarray}
\hat{\alpha}_{L}^{} & = & \hat{\alpha}_\parallel^{} - \hat{\alpha}_\perp^{}, \quad
\hat{\alpha}_{R}^{} = \hat{\alpha}_\parallel^{} + \hat{\alpha}_\perp^{}, \quad
F_V = dV - i V^2.
\end{eqnarray}

The low energy constants ($f_\pi$, $a$, $g$, $y_i$, $z_i$ $c_i$) of the Lagrangian
can be calculated by using the general ``master formula" proposed in
Ref.~\refcite{HMY06} from a class of
holographic QCD models, namely,
\begin{eqnarray}
f_{\pi}^2 & = & N_c G_{\rm YM}^{} M_{KK}^2  \int dz K_2(z)
\left[ \dot{\psi}_0^{}(z)
\right]^2, \qquad
a f_{\pi}^2  =  N_c G_{\rm YM}^{} M_{KK}^2
\lambda_1^{} \langle \psi^2_1 \rangle,
\nonumber\\
\frac{1}{g^2} & = & N_c G_{\rm YM}^{} \langle \psi_1^2 \rangle ,
\qquad
y_1^{} = {} - y_2^{} = -N_c G_{\rm YM}^{} \left\langle
\left(1 + \psi_1 - \psi_0^2 \right)^2
\right\rangle ,
\nonumber\\
y_3^{} & = & -y_4^{} = -N_c G_{\rm YM}^{} \left\langle
\psi^2_1 \left(1 + \psi_1^{} \right)^2
\right\rangle , \qquad
y_5^{} = 2 y_8^{} = -y_9^{} = -2N_c G_{\rm YM}^{}
\left\langle \psi_1^2 \psi_0^2 \right\rangle ,
\nonumber\\
y_6^{} & = & - \left( y_5^{} + y_7^{} \right) , \qquad
y_7^{} = 2N_c G_{\rm YM}^{} \left\langle \psi_1^{}
\left ( 1 + \psi_1^{} \right)
\left(1 + \psi_1^{} - \psi_0^2 \right) \right\rangle ,
\nonumber\\
z_4^{} & = & 2N_c G_{\rm YM}^{} \left\langle \psi_1^{}
\left( 1+\psi_1^{} - \psi_0^2 \right)
\right\rangle , \qquad
z_5^{} = {} - 2N_c G_{\rm YM}^{} \left\langle \psi_1^2
\left( 1 + \psi_1^{} \right) \right\rangle ,
\nonumber\\
c_1^{} & = &  \left\langle\!\!\left\langle
\dot{\psi}_0^{} \psi_1^{} \left( \frac{1}{2} \psi_0^2 + \frac{1}{6} \psi_1^2
- \frac{1}{2} \right) \right\rangle\!\!\right\rangle ,
\nonumber\\
c_2^{} & = & \left\langle\!\!\left\langle
\dot{\psi}_0^{} \psi_1^{} \left( -\frac{1}{2} \psi_0^2 + \frac{1}{6} \psi_1^2
+ \frac{1}{2} \psi_1^{} + \frac{1}{2} \right) \right\rangle\!\!\right\rangle, \qquad
c_3^{} = \left\langle\!\!\left\langle
\frac{1}{2}\dot{\psi}_0^{} \psi_1^{2} \right\rangle\!\!\right\rangle ,
\label{eq:lecshls}
\end{eqnarray}
where the smallest nonzero eigenvalue $\lambda_1^{}$ and its corresponding eigenfunction
$\psi_1^{}(z)$ satisfy the eigenvalue equation (with $n=1$)
\begin{eqnarray}
- K_1^{-1}(z)\partial_z \left[ K_2(z) \partial_z \psi_n^{} (z) \right]
&=& \lambda_n^{} \psi_n^{}(z),
\end{eqnarray}
with $K_1(z)$ and $K_2(z)$ being the warping factors of the 5-dimensional space-time.
In Eq.~(\ref{eq:lecshls}) we made use of the following definitions:
\begin{eqnarray}
\langle A \rangle & \equiv & \int_{-\infty}^{\infty}  dz K_1(z) A(z), \qquad
\langle\langle A \rangle\rangle \equiv \int_{-\infty}^\infty dz A(z).
\end{eqnarray}
Here, we take $K_1(z) = K^{-1/3}(z)$ and $K_2(z) = K(z)$ with $K(z) = 1 + z^2$
corresponding to the Sakai-Sugimoto model\cite{SS04a,SS05}.
We refer to Refs.~\refcite{MOYHLPR12,MYOH12} for the other choices.

Thanks to the master formulas, once the hQCD parameters $M_{\rm KK}$ and
$G_{\rm YM}$ are determined by, for example, the empirical values of $f_\pi$ and
$m_\rho$, all the rest of the parameters of the effective Lagrangian except the
HLS parameter $a$ are fixed .
In this work, we use
\begin{eqnarray}
f_\pi = 92.4 \mbox{ MeV}, \qquad m_\rho = 775.5 \mbox{ MeV}.
\label{PMS}
\end{eqnarray}

Armed with the master formulas and the input values, we now study the single
soliton properties.
The purpose of this study is to find the qualitatively robust features of the soliton
structure rather than quantitative agreements with nature.
Our results of soliton mass $M_{\rm sol}$, $\Delta$-$N$ mass splitting $\Delta_M$,
the winding number radius $\sqrt{\langle r^2 \rangle_W}$, and the energy radius
$\sqrt{\langle r^2 \rangle_E}$ are summarized in Table~\ref{tab:skyrmion}.
Here, we consider three versions of the HLS Lagrangian.
The full version HLS$(\pi,\rho,\omega)$ includes the pion, $\rho$, and $\omega$ mesons.
In order to see the role of the $\omega$ meson, we consider HLS$(\pi,\rho)$
that does not have th $\omega$ meson, which can be achieved by discarding the hWZ terms.
Finally the role of the $\rho$ meson can be studied by comparing the results of the model
HLS$(\pi)$ that is obtained by integrating out the $\rho$ field in HLS$(\pi,\rho)$.
For comparison, we also list the results of so-called ``the minimal model''\cite{PRV03}
that corresponds to setting $y_i = z_i = c_3 = 0$ and $c_1 = - c_2 = 2/3$ in Eq.~\eqref{eq:Lag_HLS}.
The details can be found in Ref.~\refcite{MYOH12}.

\begin{table*} [t]
\centering
\tbl{\label{tab:skyrmion}
Skyrmion mass and size obtained in models of the HLS Lagrangian.}
{\begin{tabular}{@{}ccccc@{}}
\toprule
& HLS$(\pi,\rho,\omega)$ & HLS$(\pi,\rho)$  & HLS$(\pi)$  & HLS$_{\rm min}(\pi,\rho,\omega)$   \\ \hline
\qquad $M_{\rm sol} $ (MeV) & 1184 & 834 &  922 & 1407 \\
\qquad $\Delta_M$ (MeV) & 448 & 1707 & 1014 & 259 \\ \hline
\qquad $\sqrt{\langle r^2 \rangle_W^{}}$ (fm) &  0.433 & 0.247 &  0.309 & 0.540   \\
\qquad $\sqrt{\langle r^2 \rangle_E^{}}$ (fm) &  0.608 &  0.371 &  0.417  & 0.725 \\
\botrule
\end{tabular}
}
\end{table*}

Table~\ref{tab:skyrmion} shows explicitly the role of the vector mesons in the
skyrmion properties.
The $\rho$ meson reduces the soliton mass from $922$~MeV in HLS$(\pi)$ to
$834$~MeV in HLS$(\pi,\rho)$, but the $\omega$ meson restores the large soliton
mass by giving $1184$~MeV in HLS$(\pi,\rho,\omega)$.
These tendency can also be found in the skyrmion radii.
However, the $\Delta$-$N$ mass difference, which is related to the moment of
inertia of the soliton, shows the opposite. Without the $\omega$ meson, this
mass difference becomes vary large and breaks the standard quantization method of
skyrmions. This emphasizes the important role driven by the $\omega$ meson.

\section{Dense baryonic matter from HLS}
\label{sec:matter}

The dense baryonic matter can be simulated by putting skyrmions obtained in
this model onto the FCC crystal.
For studying the baryonic matter in large $N_c$ limit, we approximate the
baryonic matter by the skyrmion matter.
The baryonic matter density $\rho$ is controlled by the crystal size $2L$ as
$\rho \propto 1/L^3$.

\begin{figure}[t]
\centerline{\includegraphics[scale=0.35]{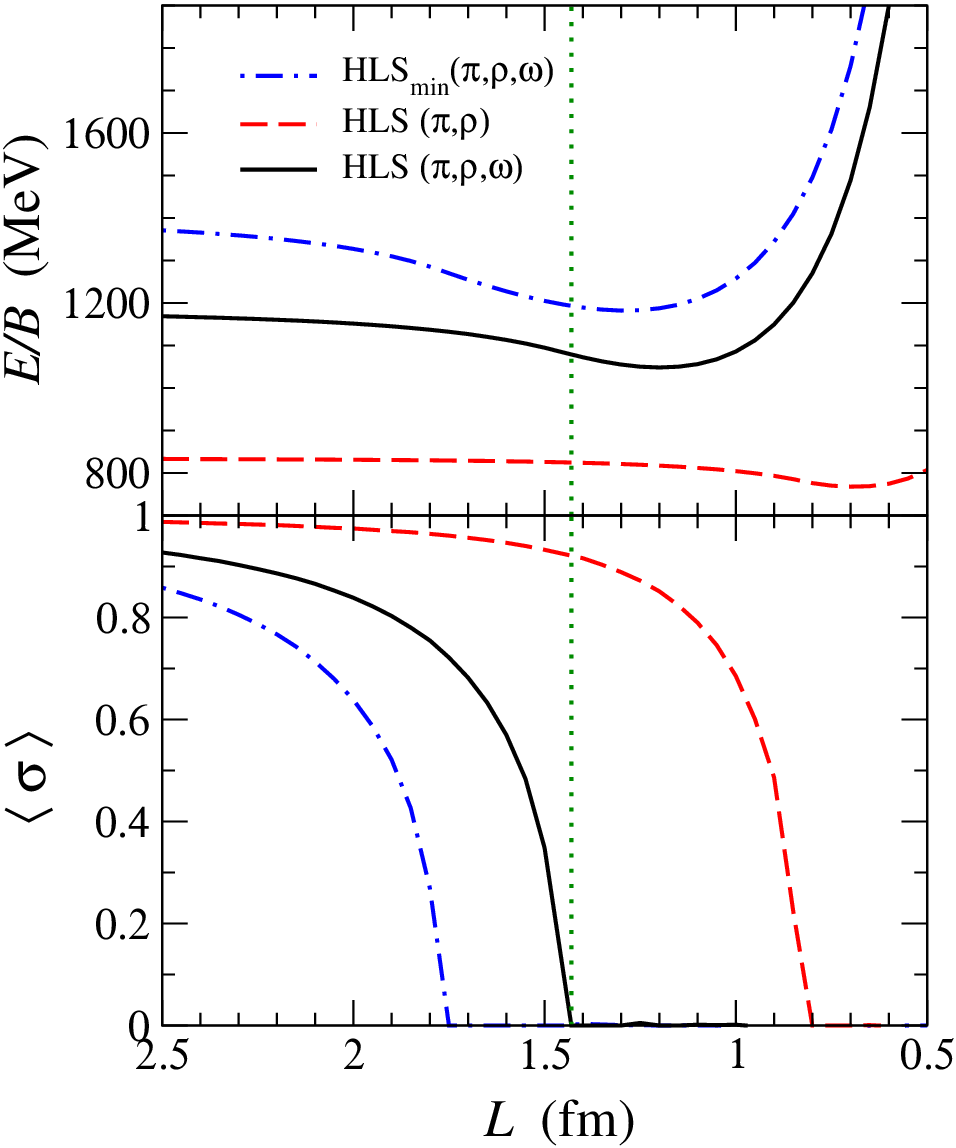}}
\caption{$E/B$ and $\langle \sigma \rangle$ of the minimum energy configuration for a given $L$.
 \label{fig:fig1}}
\end{figure}

Our results for the dependence of per-skyrmion energy $E/B$ and
$\langle \sigma\rangle$ on the FCC size parameter $L$ are shown in Fig.~\ref{fig:fig1}.
Here, $\langle \sigma \rangle$ is the space average of the chiral field $U$ over the space
that a single skyrmion occupies and it vanishes in the half-skyrmion phase.
The results for HLS$(\pi,\rho,\omega)$, HLS$(\pi,\rho)$
and HLS$_{\rm min}(\pi,\rho,\omega)$ are illustrated by solid, dashed, and
dash-dotted lines, respectively.
The position of the normal nuclear matter density is denoted by a vertical
dotted line.

Figure~\ref{fig:fig1} suggests that, in HLS$(\pi,\rho,\omega)$, $n_{1/2} \cong n_0$.
Comparing to the results of HLS$_{\rm min}(\pi,\rho,\omega)$, the $O(p^4)$ terms
and the other $\pi$-$\rho$-$\omega$ interactions through the hWZ terms make $n_{1/2}^{}$
larger by a factor $1.7$.
This higher value of $n_{1/2}^{}$ comes from the fact that the size of the single skyrmion
 is smaller in the former and that the additional interactions in HLS$(\pi, \rho,\omega)$
weaken the repulsive interactions from the $\omega$.
This conclusion is supported by the results from HLS$(\pi,\rho)$, where
$n_{1/2}^{} \cong 6n_0$.
This means that the absence of the $\omega$ meson reduces the skyrmion size to
almost half compared to that of HLS$(\pi,\rho,\omega)$.
Moreover, the weak dependence of $E/B$ on density in HLS$(\pi,\rho)$ is noticeable,
which indicates that the inclusion of the $\rho$ meson reduces the soliton mass
and almost saturates the Bogomol'nyi bound.

\section{Hadron properties with the FCC crystal background}
\label{sec:hadron}

When the meson fields in the Lagrangian is expanded in terms of the fluctuating
meson fields about their vacuum values, the meson dynamics can be investigated
in free space.
By the same way, if we incorporate the fluctuations with respect to classical solutions
for $U$, $\rho_\mu^a$, and $\omega_\mu$ discussed in the previous section,
these fields describe the corresponding mesons and their dynamics in dense
baryonic matter.
We denote the minimum energy solutions as $U_{(0)} = \xi_{(0)L}^\dagger \xi_{(0)R}^{}$,
$\rho^{a(0)}_{\mu}$, and $\omega_{\mu}^{(0)}$, and introduce the fluctuating fields
on top of the classical solutions as
\begin{eqnarray}
\xi_{L,R}^{} & = & \xi_{(0)L,R}^{} \tilde{\xi}_{L,R}^{} ,
\qquad \rho_\mu^a = \rho^{a(0)}_\mu + \tilde{\rho}^a_\mu , \qquad
\omega_\mu = \omega^{(0)}_\mu + \tilde{\omega}_\mu , \label{eq:fieldfluct}
\end{eqnarray}
where $\tilde{\xi}^\dagger_L = \tilde{\xi}_R^{} = \tilde{\xi} = \exp (i \tau_a \tilde{\pi}_a/2f_\pi)$,
$\tilde{\rho}^a_\mu$, and $\tilde{\omega}_\mu$ stand for the corresponding fluctuating fields.
By substituting these fields into the HLS Lagrangian and taking the space average
for the background field configurations as denoted by $\langle \cdot \rangle$,
we can find that the pion kinetic term is modified as
\begin{eqnarray}
\mathcal{L} & = &
\frac12 \left[ 1 - \frac23 \left( 1- \left\langle \sigma^2_{(0)} \right\rangle \right) \right]
\partial_\mu \pi^a \partial^\mu \pi^a,
\label{eq:mesoninmedium}
\end{eqnarray}
where we have taken into account only the first term of the HLS Lagrangian since
at densities away from that of chiral restoration with which we are concerned here
the other terms represent effectively higher order in the chiral counting.
The extra factor in front of the pion kinetic term \eqref{eq:mesoninmedium} suggests
that the in-medium pion decay constant $f_\pi^\ast$ reads
\begin{eqnarray}
\frac{f_\pi^\ast}{f_\pi}
=  \sqrt{1 - \frac{2}{3} \left( 1- \left\langle \sigma^2_{(0)} \right\rangle \right) } .
\label{eq:fpistar}
\end{eqnarray}

\begin{figure}[t]
\centering
\includegraphics[scale=0.29]{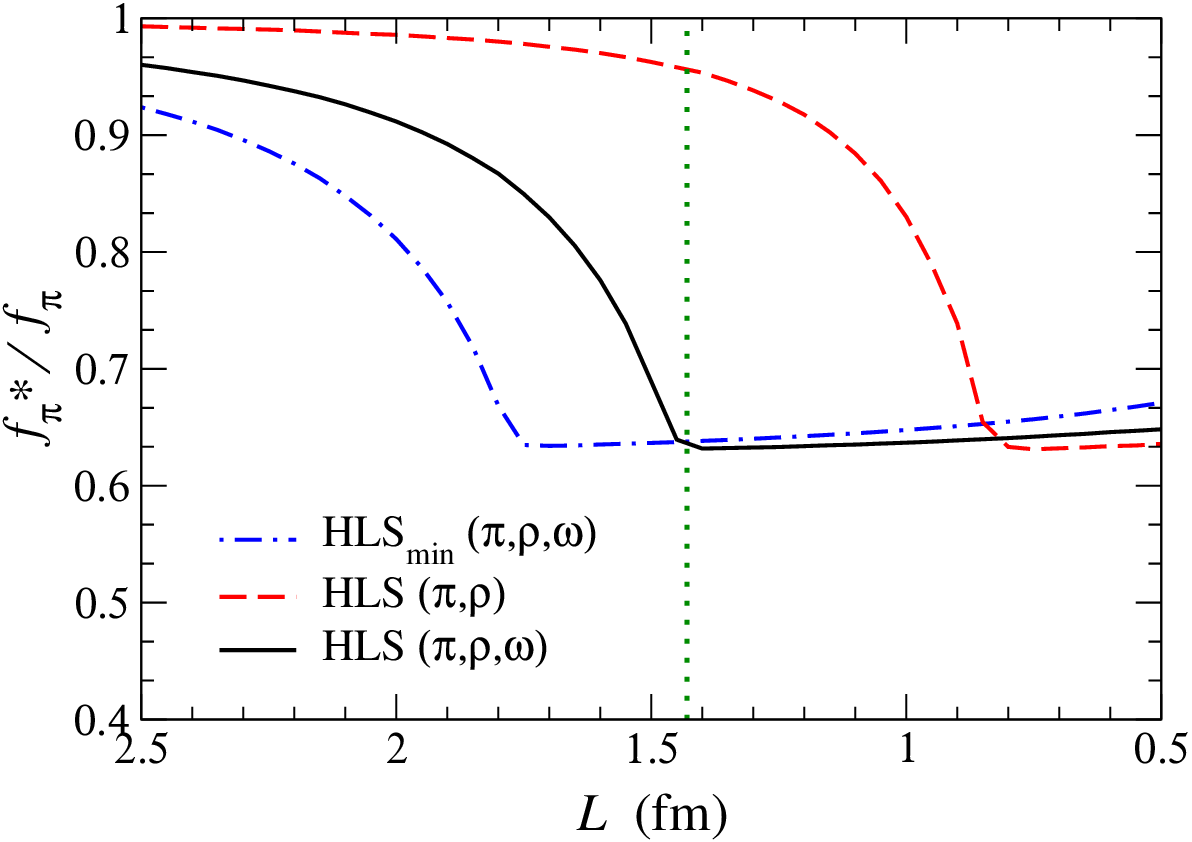} \quad
\includegraphics[scale=0.29]{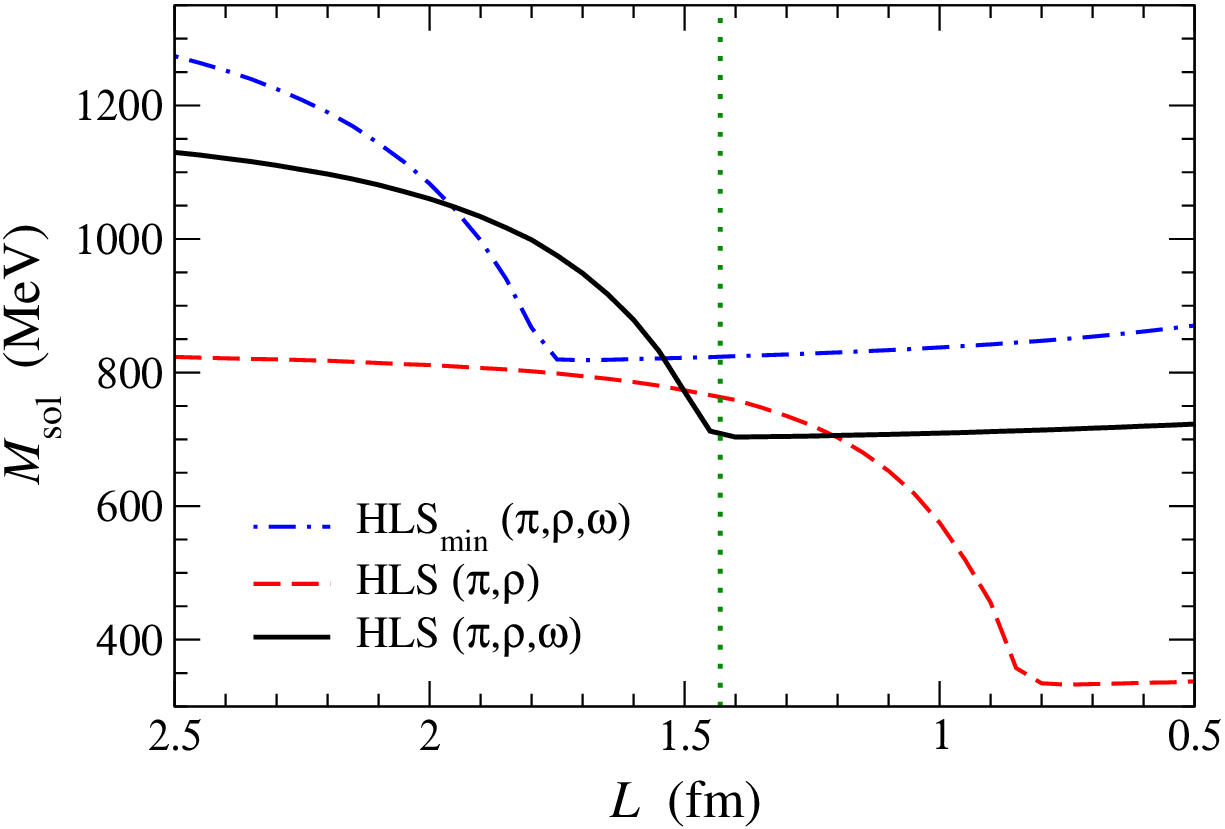}
\caption[]
{Crystal size dependence of $f_\pi^{\ast}/f_\pi$ and $M_{\rm sol}$ with
the FCC crystal background.
}
\label{fig:fig2}
\end{figure}

In the left panel of Fig.~\ref{fig:fig2}, we  present $f_\pi^{\ast}/f_\pi$ as a function
of the FCC box size.
This shows that the density dependence of $f_\pi^{\ast}/f_\pi$ in the
single-skyrmion phase is very different from that in the half-skyrmion phase.
In the single-skyrmion phase, $f_\pi^{\ast}/f_\pi$ decreases as the baryon number density
increases.
In the half-skyrmion phase, however, it almost stays at a non-vanishing value of $\sim 0.65$.
Since $f_\pi^*$ does not vanish, the half-skyrmion phase was interpreted
as a sort of pseudo-gap phase in Ref.~\refcite{LPRV03b}.

In order to determine the LECs of our effective Lagrangian we made use of
the master formulas.
As the simplest approximation in applying it to dense matter, we assume that
the in-medium modification of all the LECs can be calculated using the in-medium
values $f_\pi^{\ast}$ and $m_\rho^\ast$ which renders us to explore the medium
modified baryon mass.
With this approximation the in-medium single-skyrmion mass can be calculated and
the results are hown in the right panel of Fig.~\ref{fig:fig2}.
This shows that the soliton mass $M_{\rm sol}^{\ast}$ has a nearly identical density
dependence as $f_\pi^{\ast}$. This feature can be understood by the fact that in the skyrmion model the soliton
mass scales roughly  as $M_{\rm sol}^{\ast} \sim  f_\pi^{\ast}/e$ where $e$ is the Skyrme parameter, which follows from the scaling of the nucleon mass in the large $N_c$ limit as in matter-free space.

Our principal finding in this calculation is that the nucleon mass, which decreases
with increasing density in the skyrmion phase, stops dropping at $n_{1/2}^{}$ and
stays nearly constant during the half-skyrmion phase.
In the case of HLS$(\pi,\rho,\omega)$, the in-medium soliton mass approaches
$\sim 0.6 \, M_{\rm sol}$.
In the skyrmion phase with $n < n_{1/2}^{}$, chiral symmetry is broken with the
order parameter $\langle \bar{q}q\rangle \neq 0$ and hadrons are massive
apart from the Nambu-Goldstone bosons, i.e., the pions.
In the half-skyrmion phase, we have $\langle \bar{q}q\rangle = 0$ on the average,
but chiral symmetry is not restored yet since hadrons are still massive and there
exist pions.
There seems to be no obvious order parameter characterizing this state apart from
the presence of the half-skyrmion structure arising from a topology change.
The nucleon mass staying as a constant in the half-skyrmion phase indicates the
origin of the nucleon mass.
What this means is that the nucleon mass could have a component that remains
non-vanishing up to the chiral transition, which is reminiscent of the parity doublet
picture of baryons\cite{DK89}, where the nucleon mass is given by
\begin{eqnarray}
m_N = m_0 + \Delta(\langle \bar{q}q \rangle)
\end{eqnarray}
where $m_0$ is the chiral invariant mass and $\Delta$ is the part of
the mass that vanishes as $\langle \bar{q}q \rangle \to 0$ (in the chiral limit) for
$n \to n_c^{}$. From $M_{\rm sol}$ illustated in Fig.~\ref{fig:fig2} one can imagine the crystal size dependence of the part of the nucleon mass originning from the chiral symmetry breaking.

\section{Summary and discussions}
\label{sec:sum}

In a series of publications, we have explored to unravel the structure of baryonic matter at high density exploiting Skyrme's conjecture.
Our study includes the lowest lying vector mesons $\rho$ and $\omega$ in the 
the HLS approach with the Lagrangian given up to $O(p^4)$ in chiral counting and the leading order in $N_c$.
The LECs of the Lagrangian are then determined by using the master formulas derived
from a class of hQCD models.

Our results clearly show that the skyrmion and skyrmion matter properties are
affected by the attractive interaction of the $\rho$ meson and the repulsive interaction
induced by the $\omega$ meson.
We confirm the observation made by Sutcliffe\cite{Sutcliffe11} that the incorporation
of the tower of isovector vector mesons brings the soliton mass close to the BPS bound,
where the lowest $\rho$ vector meson plays a striking role.
As stressed by Sutcliffe, this approach to the BPS state would ameliorate the binding
energy structure of heavy nuclei, which the Skyrme model fails to describe.
What is found in this work is that the presence of the isoscalar vector meson $\omega$
obstructs this approach to the BPS state.
Whether the inclusion of the infinite tower of $\omega$'s as implied in hQCD will
change this feature is not known at the moment.

What is intriguing is that independently of how the BPS state is approached, the
Bethe-Weizs\"acker mass formula is very well reproduced by a BPS model as shown
recently in Ref.~\refcite{ANSW13}.
This model is anchored on self-dual configurations with no interactions.
The small binding energy observed is then given by a small perturbation.
This picture is diagonally opposed to the standard nuclear physics lore where the
small binding energy is given by a near complete cancellation between two large
quantities, an attractive one and a repulsive one.
Whether there is a ``dual" relation between the two opposed pictures is an intriguing issue.

On the crystal lattice, we find that the nucleon mass decreases smoothly as density
increases up to $n_{1/2}^{}$ and, in the half-skyrmion phase, the dropping of
the nucleon mass stops and the mass remains constant going toward to the
chiral restoration.
In the skyrmion phase with density $n< n_{1/2}^{}$, chiral symmetry is broken
with the order parameter $\langle \bar{q}q\rangle \neq 0$ and hadrons are
massive except the pions. However, in the half-skrymion phase, even with
$\langle \bar{q}q\rangle = 0$ on the average, chiral symmetry is not restored since
hadrons are still massive and there exist pions.
That the nucleon mass remains non-vanishing as one approaches the chiral transition
point is supported by other approaches.
As mentioned, this feature is reminiscent of the parity-doublet baryon model with a
substantially large $m_0^{}$.
Furthermore there is an indication from an extended QCD with a scalar field introduced
as an auxiliary field\cite{Kaplan13} that the structure of nucleons could be basically
different from that of mesons as in the chiral bag model\cite{Rho94}.

It should be noted that, some of our results, such as the nucleon mass, the density
$n_{\rm min}^{}$ at which per-baryon mass is at its minimum, the nucleon binding
energy and so on, deviate from nature.
This drawback indicates that in our approach some other effects, for example the
dilaton and irreducible three-body interactions, are missing.


\section*{Acknowledgments}

The work of Y.-L.M. and M.H. was supported in part by Grant-in-Aid for Scientific Research
on Innovative Areas (No. 2104) ``Quest on New Hadrons with Variety of Flavors'' from MEXT.
Y.-L.M. was supported in part by the National Science Foundation of China (NSFC) under
Grant No.~10905060.
The work of M.H. was supported in part by the Grant-in-Aid for Nagoya University Global
COE Program ``Quest for Fundamental Principles in the Universe: From Particles to the Solar
System and the Cosmos'' from MEXT, the JSPS Grant-in-Aid for Scientific Research
(S) No. 22224003 and (c) No. 24540266.
The work of H.K.L. and M.R. was partially supported by the WCU project of Korean Ministry of
Education, Science and Technology (R33-2008-000-10087-0).
Y.O. was supported in part by the Basic Science Research Program through the National
Research Foundation of Korea under Grant No.~NRF-2013R1A1A2A10007294 and
B.-Y.P. was supported by the research fund of Chungnam National University.

\end{document}